
%
\input harvmac.tex
\lref\Affrev{For a partial survey, see I. Affleck, ``Conformal Field
Theory
Approach to Quantum Impurity Problems'', UBCTP-93-25,
cond-mat/9311054.}
\lref\PG{P.G. De Gennes, {\it Scaling concepts in polymer
physics}, Cornell University Press (1979).}
\lref\AFL{N. Andrei, K. Furuya, and J. Lowenstein, Rev. Mod. Phys. 55
(1983) 331; \hfill\break
A.M. Tsvelick and P.B. Wiegmann, Adv. Phys. 32 (1983) 453.}
\lref\PF{P. Fendley, Phys. Rev. Lett. 71 (1993) 2485,
cond-mat/9304031.}
\lref\FS{P. Fendley and H. Saleur, Nucl. Phys. B388 (1992) 609,
hep-th/9204094.}
\lref\FSiii{P. Fendley and H. Saleur, ``Deriving Boundary $S$
Matrices'',
USC-94-001, hep-th/9402045.}
\lref\BN{B. Nienhuis, J. Stat. Phys. 34 (1984) 731, and references
therein.}
\lref\Smir{F. Smirnov, Phys. Lett. B275 (1992) 109.}
\lref\ZandZ{A.B. Zamolodchikov and Al.B. Zamolodchikov, Ann.  Phys.
120 (1980) 253.}
\lref\ZamoII{A.B. Zamolodchikov, Mod. Phys. Lett. A6 (1991) 1807.}
\lref\DFSZ{P. Di Francesco, H. Saleur and J.B. Zuber, J. Stat. Phys.
49 (1987)
57.}
\lref\BEG{T. Burkhardt, E. Eisenreigler and I. Guim, Nucl. Phys. B316
(1989)
559.}
\lref\BC{T. Burkhardt and J. Cardy, J.Phys. A20 (1987) L233.}
\lref\AL{I. Affleck and A. Ludwig, Nucl. Phys. B352 (1991) 849; Nucl.
Phys.
B360 (1991) 641.}
\lref\GZ{S. Ghoshal and A.B. Zamolodchikov, ``Boundary State and
Boundary $S$
Matrix in Two-Dimensional Integrable Field Theory'', RU-93-20,
hep-th/9306002.}
\lref\Car{J. Cardy, Nucl. Phys. B324 (1989) 581.}
\lref\ALii{I. Affleck and A. Ludwig, Phys. Rev. Lett. 67 (1991) 161.}
\lref\DS{B. Duplantier and H. Saleur, Phys. Rev. Lett. 57 (1986)
3179.}
\lref\GB{I. Guim and T. Burkhardt, J. Phys. A22 (1989) 1131.}

\def\t{\theta}

\def\ep{\epsilon}
\def\<{\langle}
\def\>{\rangle}
\noblackbox
\Title{\vbox{\baselineskip12pt
\hbox{USC-94-006}\hbox{cond-mat/9403095}}}
{\vbox{\centerline{Exact Theory of Polymer Adsorption}
\vskip3pt\centerline{in Analogy with the Kondo Problem}}}

\centerline{P. Fendley and H. Saleur$^*$}
\vskip2pt
\centerline{Department of Physics, University of Southern California}
\centerline{Los Angeles CA 90089}
\vskip.3in
We conjecture the exact scaling theory for the adsorption of
 two-dimensional polymers
by using boundary S matrices. We compute the boundary free energy
(the ``g-function''), study the
flow from adsorbed to desorbed phase, and derive the crossover exponent
 and all the
geometrical exponents at the transition.
 More generally, we solve
the special transition in the $O(n)$ model, the polymer case
corresponding to
$n$=0. The $n$=2 limit  appears identical to the ordinary Kondo
problem.

\bigskip
\bigskip\bigskip
\noindent $^*$ Packard Fellow
\Date{3/94}

There has recently been a surge of activity in the study of
two-dimensional
field theories with boundaries. These are of
importance in the many different
contexts of open string theory,  the Callan-Rubakov effect,
solid-state physics,
and dissipative quantum mechanics \Affrev. We point out in this letter
that
there is another interesting physical
problem in this category: the adsorption of two-dimensional polymers.
We solve
it exactly, and find a remarkable analogy with the
ordinary Kondo problem.

\bigskip

We study a long two-dimensional
polymer in
the presence of a boundary. This system can be realized experimentally
with polymeric materials that
spread reproducibly at an air-water interface \ref\VR{R. Villanove
and
F. Rondelez, Phys. Rev. Lett. 45 (1980) 1502.}. The solvent is such
that the polymer is in the universality class of lattice self-avoiding
walks.
We suppose the polymer has a short-range
interaction with the boundary, which in the lattice model corresponds
to a Boltzmann weight
$e^{\ep^s}$ per adsorbed monomer. At small $\ep^s$ entropy dominates
and the polymer is not adsorbed. As $\ep^s$ is increased, a critical
point
$\ep^s_c$ is reached where
energy and entropy terms compensate. At this point, the number of
adsorbed
monomers
varies with the total length $N$ as $N_{\rm{ads}}\propto N^{\varphi}$,
where $\varphi$ is some crossover exponent \PG. For $\ep^s\le
\ep^s_c$, the
typical size of the polymer varies with the well-known two-dimensional
bulk
exponent:
$<R^2>\propto N^{2\nu}$, $\nu=3/4$.  For $\ep^s> \ep^s_c$, energy
dominates
and at large distance the polymer behaves like a one-dimensional
object stuck
to the wall, with in particular $<R^2>\propto N^{2}$.
There is a finite fraction
of adsorbed monomers which varies as
$N_{\rm{ads}}/N\propto (\ep^s-\ep_c^s)^{-1+1/\varphi}$.

It is well known that polymers  can be described by the geometrical
$O(n)$ model
as $n\rightarrow 0$ \PG. The adsorption transition is numerically
known to occur for the
loops of the $O(n)$ model at least for
$n\in [0,1]$ (later we argue that this is true for $n\leq 2$), so
we build a
theory for it as a function of $n$.
As a surface critical phenomena this adsorption
transition can be considered as the special
transition \ref\KB{K. Binder, in {\it Phase transitions and critical
phenomena},
Vol. 8, ed. by C. Domb and J. Lebowitz, Academic Press (1983).}.
Although the
boundary is one-dimensional, it can order for geometrical models
because they
are non-local and non-unitary.
The special transition is not the only boundary critical point;
there
are also the ordinary and extraordinary transitions which correspond to
free and fixed boundary conditions in the $O(n)$
model.

Following \BN, we  consider a model of self-avoiding and
mutually-avoiding
polymer loops on some lattice
with fugacity $n$ per loop and $x_c$ per monomer. The first
non-trivial order as
 $n\to 0$ is a single
self-avoiding loop and gives the original problem. We choose the
fugacity $x_c$ so that we are at the critical point in the bulk
(this point does not depend on what happens at the boundary for
$\ep^s\leq
\ep^s_c$) and deal therefore with  very long loops.
Monomers adsorbed at the boundary have a different fugacity
$x^s\equiv
e^{\ep^s}$.
Near the boundary fixed points the problem can be viewed at large
distance
as a conformal field theory (with the
appropriate scale-invariant boundary conditions) perturbed by the
energy
operator on the boundary. At the free fixed point, this operator has
surface
scaling dimension $2$ \BC, while at the special point it has
dimension
$(m-1)/(m+1)$ \BEG, with $n=2\cos\pi/m$. The operator is relevant
at the
special point and irrelevant at the free boundary, so it induces a
flow
from special to free.

We make the crucial assumption that the boundary does not destroy
the
integrability of the scaling limit of the critical $O(n)$ model. This
is very likely since  the conformal minimal models with $\Phi_{1,3}$
perturbation on the
boundary are integrable \GZ, and the $O(n)$ models with energy
perturbation on
the boundary share many properties with these models.
We can then
require  the constraints of integrability, in particular the factorizability
of the $S$ matrix.
\bigskip

To proceed further we make an analogy with the Kondo problem of a single
species of electron ($k$=1) coupled to a spin-$1/2$
impurity. As
is well known, the $s$ waves in this three-dimensional
nonrelativistic problem reduce to a relativistic $1+1$-dimensional
problem on the
half-line, with the impurity lying on the boundary. In the following
we shall
also treat the polymers as a $1+1$-dimensional problem by performing a
Wick
rotation. It is found (for reviews, see \AFL) that there are two
critical
points in the Kondo problem. In the
UV ($T\to\infty$) the impurity is decoupled and therefore there is a
spin $1/2$
sitting
at the boundary, while in the IR ($T=0$), the impurity is screened by
a bound
electron. The model interpolates
between the two fixed points, with the Kondo temperature $T_K$ the
scale at
which the behavior crosses over from one critical point to another.
This problem is conveniently approached using $S$ matrices
\PF.\foot{We discuss
only the spin degrees of freedom; since the charge degree of freedom
does not
couple to the boundary, we ignore it.} In the bulk there is a doublet
of
left-moving particles carrying a label $1,2$ (which stands for
$S_z=\pm
\half$), and likewise for the right. The particles are massless with
dispersion
relation $E=\pm p$ which
we parametrize by $E=\mu e^{-\t}$ for the left-movers and $E=\mu
e^{\t}$ for
the right. All physics is independent of the arbitrary scale $\mu$. At
the bulk
critical point, right and left particles are independent (the two
Fermi
surfaces are infinitely far apart), so the scattering matrix
$S_{LR}=1$. The
bulk scattering matrix for two left movers with rapidities $\t_1$ and
$\t_2$
reads
\eqn\smat{S_{LL}=S_{RR}=Z(\theta)\left(I-{\theta\over\theta
-i\pi}e\right),}
where $\theta\equiv\t_1-\t_2$, $I$ is the identity matrix, and
$e=K_1$, with $$K_q=\pmatrix{0&0&0&0\cr 0&-1&q^{-1}&0\cr 0&q&-1&0\cr
0&0&0&0\cr}.$$
The function $Z(\theta)$ is a known factor ensuring unitarity and
crossing
symmetry.
The impurity can be thought of as a single particle sitting at the
boundary;
the $S$ matrix for scattering a left mover off the impurity (so it
becomes a
right mover) is
\eqn\bdrs{
\left(S_{BL}\right)_{i_1}^{j_1}=-\delta_{i_1}^{j_1}i\tanh\left
({\theta-\theta_K\over2}-{i\pi\over 4}\right),}
This is the simplest solution of the boundary ``cross-unitarity''
relation \GZ.
The ``boundary rapidity'' $\theta_K$ is related to $T_K$ by
$\theta_K\equiv\ln
(\mu /T_K)$.

The bulk $S$ matrix \smat\ for the Kondo problem
is in fact the $S$ matrix for
the critical $O(n)$ model at $n$=2 \ZamoII.  We can extend this analogy
to  the boundary case and then to other
values of $n$. Suppose we draw the trajectories of the Kondo particles
in $1+1$-dimensional space as lines in the plane.
The bulk S-matrix can be rewritten
in the manifestly $O(2)$ symmetric form as
\eqn\smatother{
\left(S_{LL}\right)_{i_1i_2}^{j_1j_2}=Z(\theta)\left[
\delta_{i_1}^{j_2}
\delta_{i_2}^{j_1}+f(\theta)\delta_{i_1i_2}\delta^{j_2j_1}\right].}
The three possible $O(n)$-invariant $S$ matrix elements are the three ways two
lines can meet each other at a vertex. The absence of
the third
invariant tensor in \smatother\ leaves only self-avoiding
trajectories, and
formally the resulting configurations are identical to the ones of the
lattice $O(2)$ model. Since the Kondo boundary scattering
is equivalent to that of a single
particle sitting at the boundary, we get in this picture a ``shadow''
line on the wall. Its effect depends on the energy scale
$T_K$;  in the UV the line is there and behaves like an adsorbed
line in the $O(2)$ lattice model, whereas
in the IR it disappears.
By this analogy we see that the Kondo flow, after appropriate
change of variables, looks very much like the flow
from special to free in the $O(2)$ model (we will make this statement
mathematically precise soon). A natural idea is then to extend the
known
solution of the Kondo problem to $n$ species and then make a
continuation to
$n=0$ to solve our polymer problem.

How to do this in the bulk is well known. The S matrix  \smatother\
 can obviously generalized to an $O(n)$ symmetric one by
allowing $i$ and $j$ to run from $1$ to $n$. We write it as
\eqn\tlconn{S_{LL}=Z(\theta)\left[I+f(\t) e^{(n)}\right],}
The function $f$ is determined by requiring that the $S$ matrix obeys
the
Yang-Baxter equation. The matrix $e$ in \smat\ satisfies the
Temperley-Lieb algebra relations $TL(2)$ at the particular value $n$=2
of the
usual parameter, while the $e^{(n)}$ matrices provide a representation
of the
$TL(n)$ algebra. Using this algebra, one finds that the Yang-Baxter
equation
reduces solely to a functional equation for $f(\t)$ depending on $n$.
The
solution is \ZamoII
\eqn\selts{f(\theta)={\sinh(\theta/m)\over
\sinh (\theta-i\pi)/m},}
where $n=q+q^{-1}$ and $q=\exp{i\pi\over m}$.
Physical predictions depend only on the algebra satisfied by
the $S$ matrices and not the particular representation of the algebra
\refs{\ZamoII,\Smir,\FS}. We can thus use the other representation
$e^{(n)}=K_q$.
Because this representation makes sense for $n$ non-integer, it provides
the desired analytic continuation.
The boundary $S$ matrix is \bdrs\
in all of these representations.

To conclude: we conjecture that the bulk S matrix \tlconn\ with
$e^{(n)}=K_q$ and
the boundary $S$ matrix \bdrs\ describe the full field theory
that interpolates between special and free boundary conditions for the
$O(n)$ model with $n\le 2$. This theory has massless bulk particles
and a boundary $S$ matrix depending on a scale $T_K$.

\bigskip

To test this conjecture, we show that it gives a number of
quantitative
predictions for the special transition, all consistent with known
results. We use the thermodynamic
 Bethe ansatz (TBA) to derive the free energy resulting from the
boundary $S$
matrix.  As usual, instead of looking at right and left movers on the
half-line, we look at the equivalent problem of left movers on the
full line.
This transforms the boundary
into a ``particle'' fixed at the origin. To quantize the momenta, instead
of the
full line we take space to be a circle of length $L$. We therefore
consider
the $O(n)$ model as a $1+1$-dimensional problem
at quantum temperature $T$. Physically  this
 corresponds to the
statistical mechanical problem on a torus  of length $L$ and
circumference
 $R={1\over T}$, with one impurity line in the middle.  There is a
bulk free energy independent of the
boundary coupling, as well as a surface or impurity free energy.
To compute the
latter we need to specify the value of $n$, and use a representation
of the
$S$ matrix algebra for this particular value. The simplest approach will
be to make computations for $n=2\cos(\pi/m)$ where $m$ is an integer,
and
then to continue naively the results to $m$ non-integer or $m=2$. The
bulk calculation can be found, for example, by taking the zero-mass
limit of
the $O(n)$ model calculation in \FS.
The free energy is given in terms of the pseudo-energies
$\epsilon_j(\theta)$
obeying
the set of integral equations
\eqn\inteqs{\epsilon_j(\theta)=\delta_{j1}e^{-\theta}-
\int{d\theta'\over 2\pi}
{1\over\cosh(\theta-\theta')}\left(\ln\left(1+e^{-\epsilon_{j-1}
(\theta')}\right)
+\ln\left(1+e^{-\epsilon_{j+1}(\theta')}\right)\right),}
where $j=1,\ldots,m-2$ and $\ep_0=\ep_{m-1}\equiv \infty$. Including a
non-trivial
boundary $S$ matrix changes the quantization condition for the
particles'
momenta, which changes the density of states. This in turn adds an
extra piece
to the free energy, yielding
\eqn\fimp{f_{\rm{imp}}=-T\int{d\theta\over
2\pi}{1\over\cosh(\theta-\ln(T/T_K))}
\ln(1+e^{-\epsilon_1(\t)}).}
The Kondo result \AFL\ is given by taking $m\to\infty$.

Let us now extract physical predictions from this. The easiest result
to get is
the cross-over exponent $\varphi$. To obtain it we observe,
following \ref\ZamoI{Al.B. Zamolodchikov, Nucl. Phys. B358 (1991),
524.} that the system \inteqs\ implies the periodicity of
the
pseudo-energies: $\epsilon_j[\theta+(m+1)i\pi]=\epsilon_j(\theta)$. As
a consequence, close to $\theta=\infty$
we can expand
\eqn\expan{
\ln(1+e^{-\epsilon_j(\theta)})=\sum_{k=0}^\infty
 L_j^{(k)}\left(e^{-2\theta/(m+1)}
\right)^k.}
Moreover one can show by explicitly plugging the expansion into the
equations
\inteqs\ that the term $k=1$
vanishes identically. The $k=0$ term follows from the solution of the
system
\eqn\syst{x_j^2=\left(1+x_{j-1}\right)
\left(1+x_{j+1}\right),}
where $x_j\equiv e^{-\epsilon_j(\infty)}$, yielding
\eqn\sol{1+x_j=\left[{\sin\pi(j+1)/(m+1)\over\sin\pi/(m+1)}\right]^2.}
The expansion \expan\ enables us to find the UV behaviour of
$f_{\rm{imp}}$. As
$T\rightarrow\infty$, the integral is dominated by the $\theta$ large
region
where the expansion is expected to be valid, so that
\eqn\uvf{{f_{\rm{imp}}\over T}=-{1\over
2}\ln(1+x_1)+\sum_{k=2}^\infty
 \left({T_K\over T}\right)^{2k/(m+1)}f_{UV}^{(k)}.}
plus perhaps some non-universal bulk terms. Thus we recover the fact
that the
dimension of the energy operator at the special transition is
$1-2/(m+1)$.
Standard finite-size and boundary scaling arguments imply that
$f_{\rm{imp}}$ is a function of $(x^s-x_c^s)^{\nu/\varphi}R$ where
$\nu$
is the usual \BN\ thermal exponent, $\nu={m+1\over 4}$. Also, from
perturbation
theory we
expect that $f_{\rm{imp}}$ is analytic in $x^s-x_c^s$. Therefore
$$
{\varphi\over \nu}={2\over m+1}
$$
and $\varphi=1/2$ as desired \BEG. Moreover we deduce from the
absence of the $k=1$ term in the expansion \uvf\ that the
one-point
function of the boundary energy vanishes at the UV fixed point,
a known characteristic of the special transition \BC.
In the $m\to\infty$ Kondo limit, the expansion results in
the
familiar log terms of the Kondo problem.

We also find a similar expansion near the free fixed point
($T\rightarrow 0$). In this limit only the region
$\theta$ negative and large contributes to the impurity free energy.
One
can thus expand out the $1/\cosh$ in \fimp\ as
$${1\over 2\cosh(\t-\ln(T/T_K))}\approx {T\over T_K}
e^{-\t}-\left({T\over
T_K}\right)^3 e^{-3\t}+ \dots.$$
We can integrate each piece individually because $\ep_1 \approx
\exp(-\t)$ as
$\t\to-\infty$, yielding
\eqn\irf{{f_{\rm{imp}}\over T}=\sum_{k=1}^\infty
 \left({T\over T_K}\right)^{2k-1}f_{IR}^{(k)}.}
Notice that it goes to zero at $T=0$. The power of the first
correction
indicates that the dimension of the energy operator at the free fixed
point is
$1-(-1)=2$, as expected \BC. Moreover, by using perturbed conformal
field
theory, one in fact finds that the universal contribution to the $T^2$
term
must vanish, as indeed seen in \uvf.
\bigskip

The crucial identification with the special transition can be obtained
by
looking at the evolution of the boundary ``$g$-function''
\refs{\Car,\ALii}. On
a long strip of length $L$, the partition function with boundary
conditions $A$
and $B$ at the sides reads  $Z=g_A g_B \exp(-E_0L)$. Thus $g$
associated with our boundary condition should be
$g=\exp(-f_{\rm{imp}}/T)$.
There are a number of subtleties which arise \FSiii; however, for
massless
theories these should result only in multiplying $g$ by an overall
constant
independent of scale. Thus from the TBA we see how $g$ flows from the
UV
(special) to IR (free), the expansions \uvf\ and \irf\ giving
\eqn\gdif{{g^{UV}\over g^{IR}}=\left(1+x_1\right)^{1/2}=
 {\sin 2\pi/(m+1)\over\sin\pi/(m+1)}.}

To derive $g^{UV}$ and $g^{IR}$ from conformal field theory, one needs
to
construct the boundary states in the manner of \Car. In the $O(n)$
model, this
is problematic because of the non-locality; we discuss this below.
However,
without knowing these states explicitly we can still
find the boundary ``fusion'' operator $\Phi$ which changes the
boundary
conditions from free to special. This enables us to determine the
ratio
$g^{UV}/g^{IR}$, and confirm the result numerically. If the boundary
conditions
$A$ and $B$ are such that only the state created by the operator
$\Phi_{r,s}$
(and its descendants) propagate along the strip, it follows that for
minimal
models \refs{\Car,\ALii} that $g_Ag_B= S^{r,s}_{1,1}$, where
$S^{r,s}_{r',s'}$
is the modular-transformation matrix for the minimal-model characters.
We
choose $A$ to be free boundary conditions, and $B$ such that
$\Phi_{r,1}$
propagates. We then change the boundary condition $A$ to the special
boundary
conditions by inserting the boundary operator $\Phi$. If
$\Phi=\Phi_{1,s}$ for
some $s$ then
\eqn\gdifii{{g^{UV}\over g^{IR}}={S^{r,s}_{1,1}\over S^{r,1}_{1,1}}
= {\sin s\pi/(m+1)\over\sin\pi/(m+1)}.}
Comparing with \gdif, we see that the TBA indicates that
$\Phi=\Phi_{1,2}$.
This is consistent with the Kondo case,  where $\Phi$ is the $SU(2)_1$
primary
field  \AL. This field has bulk dimension $1/4$, which is the $m\to\infty$
limit of
the dimension of $\Phi_{1,2}$ field.

To check this numerically, first recall that the surface exponents for
the
``fuseau'' operators  (operators which force $L$ lines down the strip)
are
$d^s_{L}=h_{L+1,1}$ \DS.
It follows that if both $A$ and $B$ are free boundary conditions and
$L$
lines go along the strip, the state that propagates corresponds to the
operator
$\Phi_{L+1,1}$. Thus if we change the boundary conditions on both
sides of the
strip from free to special, the operators in the OPE $\Phi\times\Phi\times
\Phi_{L+1,1}$ are those which propagate. For $\Phi=\Phi_{1,2}$, this
means that
$\Phi_{L+1,3}+\Phi_{L+1,1}$ propagate.
It is easy to see that $h_{L+1,3}<h_{L+1,1}$ for $L\geq 1$. Hence the
exponent
of the $L$-lines fuseau operator
at the special point should be $h_{L+1,3}$ while it is $h_{L+1,1}$ at
the free
point. It is easy to investigate this question numerically by
explicitly
diagonalizing the transfer matrix of the lattice $O(n)$ model. The
question of
surface exponents has
been addressed in \refs{\DS,\GB}. In
\GB\ the authors study the adsorption transition
for polymers but unfortunately they do not discuss the fuseau
operators for
$L>1$ nor the case $n\neq 0$. We have therefore extended their
analysis
to these cases. The results look as usual and confirm completely that
$\Phi_{L+1,3}$ propagates down the strip at the special point, giving
strong
evidence indeed for our conjecture.
We give some examples for $n=1$ below.

\bigskip

Exponents for polymers follow from the above by choosing $m=2$. Other
quantities
can be obtained by studying values near $m=2$  and
taking appropriate
derivatives. For instance  the ratio \gdifii\ is of course equal to
one
for $n=0$ but the result for one polymer loop follows from the first
non-trivial term:
 $g^{UV}/g^{IR}=2/3\sqrt{3}$.  The analysis of the free energy
 for non-integer values of $m$
is technically more complicated \FS, but for the polymer one can obtain it as
an analytic function of $R^{2/3}
(x^s-x_c^s)$. One expects generally that it is proportional to the number of
 adsorbed monomers $N_{\rm{ads}}$, and that
\eqn\adsscal{N_{\rm{ads}}=N^\varphi F[(x^s-x_c^s)N^\varphi],}
where $F$ is a scaling function that would be most interesting to
compute.
What we obtained of course looks qualitatively like this, but instead
of $N$ we have the variable $R$. By finite-size scaling, the typical
length of the polymer on a cylinder of radius $R$ at critical bulk
coupling varies as $N\propto R^{4/3}$, so using $\varphi=1/2$ our
expansion
\uvf\ looks indeed like \adsscal.

Equation \gdif\ can also be continued to $n<0$.
In this regime, $g^{UV}<g^{IR}$, so the ``$g$-conjecture'' \ALii\
does not hold for non-unitary models, as was the case for the
$c$-theorem of
Zamolodchikov. Another interesting characteristic is that we find
perfectly reasonable
results at $n\ge 1$ too. The
standard result \ref\BT{R. Balian and G.Toulouse, Ann. Phys. 83 (1974)
28.} that
the $O(n)$ model has a one-dimensional phase transition only for
$n\leq 1$
can naively be used to argue that the
special transition disappears for $n>1$. Think, however, of the
lattice $O(n)$
model. The high temperature
expansion of the Ising model gives an $O(n=1)$ model with a
fugacity $x=\rm{th}\beta$
per monomer. The special transition in the loop model occurs at the
value
of fugacity for monomers at the boundary that corresponds
to $\beta=\infty$, since the Ising model in one dimension
orders only at zero temperature: $x^s_c=1$. Thus there {\it is} a
special
transition in the
geometrical
$O(n)$ model with $n$=1 although it is usually agreed that there is
none
for the Ising model. The Ising spin model is a subset of
the geometrical model
only. The geometrical objects
experience a transition, but the geometrical properties do not have
a local meaning in terms of spins.
For $n>$1, it is possible that there is a finite transition for a
finite
boundary monomer fugacity which however would correspond to sending
the boundary coupling of the $O(n)$ model into
the complex plane. If so, one really could consider
the spin degrees of freedom in the Kondo problem
as the $n$=2 limit of the special transition of the $O(n)$ model.
This certainly seems
true from the formal point of view of the $S$ matrices and the TBA.

Measurements of the surface exponents at the special transition in the
$n$=1
model are easier at this point because we know $x_c^s$ exactly;
for other
values of $n$ it
must be determined numerically.  We thus numerically diagonalize the
transfer
matrix at $n$=1 and find the gaps in the spectrum, which are
proportional to
the scaling dimensions. With special boundary
conditions on both sides of a strip of width $M$ lattice sites,
these
dimensions are
\ifx\answ\bigans\vfill\eject\fi
\medskip
\settabs 7 \columns
\+&$M$=3&4&5&6&$\infty$&\rm{expected}\cr
\+$L$=2&.19106&.17961&.17340&.16923&.167(2)&1/6\cr
\+$L$=3&1.19444&1.13747&1.10741&1.08955&1.02(2)&1\cr
\+$L$=4&2.03612&2.75960&2.74731&2.72118&2.6(2)&5/2\cr
\medskip
\noindent
This confirms our identification of the $\Phi_{1,2}$ operator as the
boundary
fusion operator.
We also find from the above results
$g^{UV}/ g^{IR}=\sqrt{2}.$
This cannot really be interpreted in terms of spin.\foot{We do note,
however,
that up to a constant shift the free energy at $n$=1 should also describe the
flow from
free to fixed boundary conditions in the Ising model \ALii, which
has
precisely this ratio. This case corresponds to perturbation by the
magnetic
field, which is not invariant under spin flip and is not equivalent to
our problem. However, one can check the dimensions of the magnetic
field at the
free and fixed points, and one finds that they are the same as those
of the
energy operator at the special and free points, respectively.}
Indeed, the IR
fixed point
is just free boundary conditions for the high temperature contours
and
therefore also for the Ising spins, hence $g^{IR}=1$ \Car. The UV
fixed point
corresponds to infinite coupling of Ising spins at the boundary, so
these spins can be either all  $+$ or  all $-$. Moreover, summing
over
these two choices leads essentially to free boundary conditions for
the spins
next to the boundary. Hence the UV fixed point should be interpreted
as some
superposition
of (two possible) fixed and one free boundary condition. This can
be made more quantitative by recalling that the one-point function of
the
energy operator vanishes at the special transition.
This happens for a boundary state invariant under duality, for
example
the superposition
$\half( |\rm{fixed }+\>+ |\rm{fixed }-\>+\sqrt{2}|\rm{free}\>)$.
There is no such boundary state in the
Ising spin model since the coefficients are not integers. However,
there
can be such a state in the non-minimal loop model where the condition
of
integrality
is relaxed \DFSZ.
\bigskip
We have concentrated on the flow towards the desorbed phase, but the
flow
towards the adsorbed one should not be very different since once a
polymer
is adsorbed at the boundary and has fractal dimension one, the other
ones see
free boundary conditions on  a slightly smaller system. We have
checked numerically that in the adsorbed phase the fuseau exponents
 again have the value $d_L^s=h_{L+1,1}$.

\bigskip
\noindent
{\bf Acknowledgments}: This work was supported by the Packard
Foundation, the
National Young Investigator program (NSF-PHY-9357207) and  the DOE
(DE-FG03-84ER40168). We thank P. Dorey for useful
correspondence.

\listrefs

\bye